\documentclass[12pt]{article}

\usepackage{sbc-template}
\usepackage{graphicx,url}
\usepackage[utf8]{inputenc}
\usepackage[brazil]{babel}

\title{Impacto de Treinamentos em Programação Competitiva no Ensino Médio: Resultados e Desafios}

\author{Camila da Cruz Santos\inst{1}, Sarah Souto dos Santos\inst{1}, Crishna Irion\inst{1}, \\ Giullia Rodrigues de Menezes\inst{1},
Rafael Dias Araújo\inst{1}, João Henrique de Souza Pereira\inst{1}}

\address{Faculdade de Computação (FACOM) -- Universidade Federal de Uberlândia
  (UFU)\\
  Uberlândia -- MG -- Brasil
  \email{ \{camilacruz, sarah.santos, crishna \}@ufu.br } \email{ \{giullia.rodrigues, rafael.araujo, joaohs\}@ufu.br }
}

\begin{document} 

\maketitle

\begin{abstract}
This article presents an ongoing research aiming to develop an effective methodology for teaching programming, focusing on participation in the Brazilian Informatics Olympiad (OBI), for elementary and high school students. The training conducted with students from the Federal Institute  and state schools, demonstrates the importance of programming training programs as a way to promote interest in computing, stimulate the development of computational skills, and increase participation in competitions such as the OBI.
The next steps of the research include conducting more training cycles and analyzing the results obtained in the competitions.
\end{abstract}
     
\begin{resumo} 
  Este artigo apresenta uma pesquisa em andamento que visa desenvolver uma metodologia eficaz para o ensino de programação, com foco na participação na Olimpíada Brasileira de Informática (OBI), para alunos do ensino fundamental e médio. O treinamento realizado com alunos do Instituto Federal e escolas estaduais, demonstra a importância de programas de treinamento em programação como forma de promover o interesse pela computação, estimular o desenvolvimento de habilidades computacionais e aumentar a participação em competições como a OBI. 
  As próximas etapas da pesquisa incluem a realização de mais ciclos de treinamento e a análise dos resultados obtidos nas competições.
\end{resumo}

\section{Introdução}

A Base Nacional Comum Curricular (BNCC), homologada em 2022 pelo MEC, estabeleceu o Pensamento Computacional (PC) como um eixo essencial na educação básica brasileira \cite{MECparecer}. O PC é definido como a habilidade de identificar e aplicar conceitos de computação para resolver problemas cotidianos, integrando-se de forma interdisciplinar ao currículo escolar e promovendo habilidades essenciais para o século XXI \cite{wing2006}.

Uma maneira de abordar os objetivos de aprendizagem da BNCC é por meio da Aprendizagem Baseada em Problemas - ABP, que busca atualizar conhecimentos, qualificar profissionais, desenvolver habilidades e ensinar a lidar com novas tecnologias \cite{Martins2002phd}. Nesse cenário, a programação competitiva, na qual os competidores desenvolvem programas para resolver desafios computacionais, pode ser empregada como uma maneira de implementar a Aprendizagem Baseada em Problemas\cite{Laaksonen2017}. 


A principal competição de educação básica no Brasil é a Olimpíada Brasileira de Informática (OBI), organizada pela Sociedade Brasileira de Computação (SBC), e será o objeto de estudo desta pesquisa \cite{sbcOBI}. 
O Instituto Federal do Triângulo Mineiro (IFTM) foi escolhido como ambiente para estudo de caso deste trabalho, ao considerar os cursos ofertados e dada a relevância para a região.

Este trabalho tem como objetivo principal mostrar a importância de treinamento em programação como forma de promover o interesse pela computação, estimular o desenvolvimento de habilidades computacionais e aumentar a participação em competições científicas. 
Este trabalho contribui com a comunidade de Educação em Computação e Informática na Educação ao propor treinamentos de programação voltados à programação competitiva, levando a perspectiva de aumentar o interesse pela tecnologia e a computação, estimular o desenvolvimento de habilidades computacionais e aumentar a participação em programação competitiva. 

O artigo está organizado como descrito a seguir. A Seção \ref{sec:trabalhos} apresenta uma visão geral de trabalhos correlatos encontrados na literatura. A Seção 3 apresenta a correlação entre a OBI e o Pensamento Computacional. Posteriormente, na Seção \ref{sec:metodologia}, é descrita a metodologia utilizada neste trabalho, seguidos dos resultados (Seção \ref{sec:resultados}), bem como a análise dos dados apresentados. A seguir, são apresentados os próximos passos da pesquisa (Seção \ref{sec:passos}) e por fim, as considerações finais são expostas na Seção \ref{sec:conclusao}.

\section{Trabalhos Relacionados} \label{sec:trabalhos}


Piekarski et al. (2023) \nocite{PIEKARSKI2023} implementou um programa de extensão para lecionar programação e permitir a participação em competições de programação para alunos de ensino técnico em uma escola estadual de 2016 a 2022. No texto os escritores descrevem as ferramentas empregadas e as fases do treinamento conduzido. A capacitação envolveu 20 horas de instrução e prática no Beecrowd.


Vitorino et al. (2018) \nocite{SBIE2018vitorino} desenvolveu um trabalho analisando medalhistas que participaram das olimpíadas de informática, coletando dados de entrevistas e pesquisas. O objetivo era entender as circunstâncias que estimulam os participantes, bem como as práticas utilizadas para obter o melhor desempenho. Esta pesquisa observou que os premiados em categorias avançadas utilizam ferramentas com juízes online mais do que aqueles em categorias de iniciantes.

Sousa et al. (2021) \nocite{wei} apresentou em seu relato de experiência a preparação de alunos de graduação em Tecnologia da Informação (TI) da Universidade Federal do Ceará (UFC) para a Olimpíada Brasileira de Informática (OBI) Nível Sênior e concluiu, após implantar um programa de treinamento para programação competitiva, com preparação, criação de uma plataforma de conteúdos e exercícios que a importância da aplicação de uma metodologia de treinamentos é fundamental para a formação dos alunos e para a consolidação da cultura de programação competitiva.

\section{A OBI e a promoção do Pensamento Computacional}

Competições de programação são competições que desafiam os participantes a resolver problemas de ciência da computação em um tempo limitado, escrevendo códigos que devem ser executados corretamente. Envolvem conceitos principais como a compreensão e aplicação de algoritmos para resolver problemas que podem incluir lógica, matemática e estruturas de dados; entre outras \cite{Lertbanjongam2022}.

A Olimpíada Brasileira de Informática (OBI) é uma competição de programação organizada anualmente, desde 1999, pela Sociedade Brasileira de Computação (SBC). A OBI objetiva despertar o interesse dos alunos pela Computação e pelas Ciências em geral, além de fomentar a introdução do raciocínio computacional e das técnicas de programação nas escolas de ensino fundamental e médio \cite{sbcOBI}.

A prova é dividida em duas modalidades, de acordo com a fase escolar dos participantes: (i) iniciação: destinada a alunos do quarto ao nono ano do ensino fundamental, esta modalidade foca em problemas de lógica que envolvem conhecimentos prévios para o projeto de algoritmos; (ii) programação: aberta a alunos de qualquer série do ensino fundamental e médio, estudantes do quarto ano de cursos técnicos de nível médio e alunos que estejam cursando pela primeira vez o primeiro ano de um curso superior. Esta modalidade exige a implementação de soluções computacionais para os problemas propostos.

Ambas as modalidades são organizadas em três fases: Local, Estadual e Nacional. Os melhores participantes em suas respectivas modalidades e níveis avançam para a fase seguinte \cite{sbcOBI}. As duas são separadas por níveis, a modalidade Programação, foco deste trabalho, possui quatro níveis: 
\begin{itemize}
    \item Nível Júnior (PJ): para alunos até o nono ano do ensino fundamental;
    \item Nível 1 (P1):  alunos do ensino fundamental até  o primeiro ano do ensino médio;
    \item Nível 2 (P2): alunos do ensino fundamental até o terceiro ano do ensino médio;
    \item Nível Sênior (PS): alunos que estejam cursando o quarto ano de escolas do Ensino Técnico ou que estejam cursando, pela primeira vez, o primeiro ano de um curso de graduação.
\end{itemize}

Quando participam da OBI, os estudantes precisam resolver desafios que requerem a elaboração de algoritmos complexos, sem consulta ou auxílio externo. Para isso, é necessário que eles consigam dividir problemas complexos em tarefas menores, identificar padrões e abstrair os dados de maneira correta. O que favorece o desenvolvimento do pensamento computacional. Além disso, a obrigação de testar e depurar softwares auxilia na aquisição de um entendimento completo do procedimento de desenvolvimento, bem como na capacidade de identificar e corrigir falhas de forma minuciosa.

O treinamento para OBI é um exemplo de ferramenta que segue a abordagem ABP, que destaca o aluno como o centro do aprendizado, motivando-o a enfrentar desafios reais e complexos. A competição apresenta vários desafios ideais para aplicar a ABP. Durante a OBI, os estudantes são treinados na prática e desenvolvem habilidades para lidar com desafios complexos e sem uma estrutura definida. Por fim, participar da OBI incentiva a autonomia dos estudantes, que são encorajados a buscar soluções por si mesmos, mesmo recebendo apoio de materiais de ensino e supervisores.

\section{Metodologia} \label{sec:metodologia}
Para atingir o objetivo da pesquisa de desenvolver uma metodologia eficaz de estudo e treinamento voltada para os estudantes do ensino fundamental e médio, além de tornar acessível o conhecimento de programação e participação dos alunos na OBI, esta pesquisa adotou uma abordagem híbrida, combinando metodologias de pesquisa qualitativas e quantitativas. Este trabalho foi baseado no treinamento proposto por \cite{Menezes2024}.


\subsection{Etapas da pesquisa}

\begin{figure}[htb!]
     \centering
     \vspace{-0.5cm}
     \includegraphics[width=0.7\linewidth]{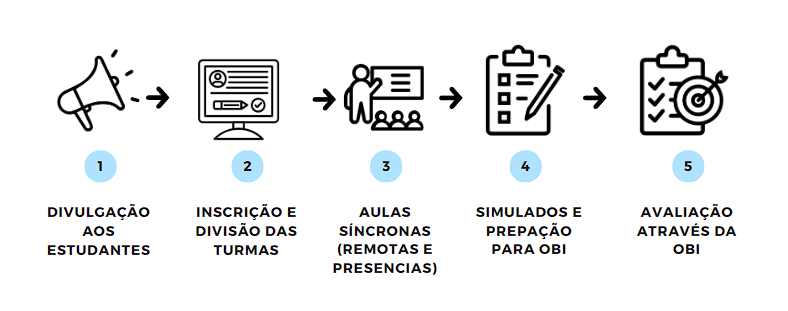}
     \vspace{-0.6cm}
     \caption{Fluxo de atividades realizadas durante o treinamento}
     \label{fig:fluxo}
 \end{figure}

A pesquisa foi estruturada em cinco fases principais, sintetizadas na Figura \ref{fig:fluxo}. Primeiro, as pesquisadoras e professores parceiros divulgaram o treinamento entre os estudantes, utilizando campanhas presenciais e redes sociais para atrair interessados em programação. Em seguida, os alunos que demonstraram interesse inscreveram-se voluntariamente, e as turmas foram organizadas conforme a disponibilidade de horários.

O ensino foi então conduzido de forma síncrona, tanto presencial quanto à distância, utilizando ferramentas como \textit{Neps Academy} e \textit{Beecrowd Academic}. Durante o treinamento, a equipe auxiliou os estudantes na resolução de problemas e na criação de algoritmos. Com a aproximação da OBI, a preparação se intensificou, com foco em exercícios e simulados específicos para capacitar os alunos aos desafios da competição. Finalmente, os estudantes participaram da prova da OBI, permitindo a análise da eficácia do treinamento oferecido.

A equipe executora é composta pelas pesquisadoras deste estudo, que atuam como instrutores dos conteúdos. Além de alunos da graduação dos cursos de Sistemas de Informação e Ciência da Computação da UFU, que fazem parte da disciplina projeto de extensão e desempenham a função de monitores no treinamento, tanto na modalidade presencial quanto remota. As atividades realizadas estão descritas na tabela \ref{tab: cronograma}, assim como os conteúdos abordados em cada etapa e as ferramentas utilizadas.

\begin{table}[htb!]
    \centering
    \caption{Cronograma de aulas e ferramentas utilizadas}
    \label{tab: cronograma}
    \includegraphics[width=0.7\textwidth]{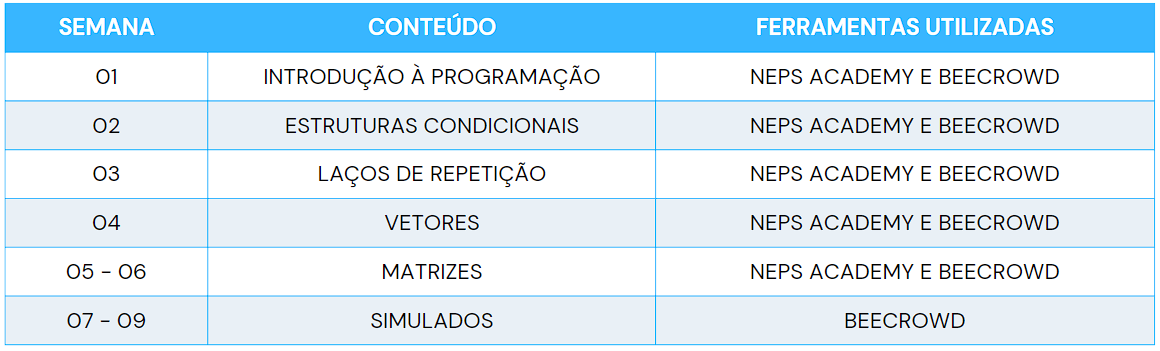}
\end{table}

\subsection{Formação  das Turmas}

A divulgação das turmas para o treinamento ocorreu de forma presencial nas escolas e através das redes sociais, resultado da cooperação entre os pesquisadores e os professores parceiros. Isso possibilitou que os estudantes interessados se matriculassem e participassem, não importando o nível de familiaridade que tinham com programação anteriormente. A decisão de não realizar uma seleção baseada em habilidades prévias estava de acordo com o propósito do projeto para essa primeira fase: disseminar conhecimento em programação e estimular a participação na OBI.

O treinamento é oferecido nas modalidades presencial e  remota. A modalidade presencial é realizada no Campus Uberlândia Centro, onde foram ofertadas 90 vagas, distribuídas em  três turmas. As aulas são realizadas três vezes por semana, para alunos do nono ano do ensino fundamental e do primeiro ano do ensino médio, com duração média de duas horas, ao longo de dois meses e meio. Na modalidade síncrona remota, por conseguinte, foram disponibilizadas 80 vagas para alunos do ensino médio exclusivamente do IFTM. As aulas ocorrem aos sábados seguindo a mesma duração. 


\subsection{Ferramentas Utilizadas}

Os estudantes participantes da turma remota foram organizados em uma sala de aula virtual utilizando o software \textit{Microsoft Teams} e \textit{Discord}, onde estão disponíveis todas as atividades, aulas e material complementar. Através da plataforma os estudantes possuem espaço para tirar as dúvidas utilizando salas de videoconferência e fórum. 


Para gerenciamento das atividades, a plataforma \textit{Neps Academy} foi selecionada, pois em um estudo sobre os vencedores da OBI 2019 feito por \cite{Menezes2021}, a plataforma \textit{Neps Academy} foi identificada como a preferida pelos alunos,  devido ao seu ambiente propício para o aprendizado. Adicionalmente, o \textit{Beecrowd Academic} é empregado como um recurso adicional, possibilitando a elaboração de disciplinas e testes práticos, o que simplifica a monitorização do progresso dos participantes.

O ensino de programação é feito através de trilhas de estudo em C++, com ênfase na OBI, criada pela plataforma \textit{Neps Academy}. Embora haja aulas síncronas, a proposta é que o professor guie o estudante na trilha, mas cada um pode progredir no conteúdo de acordo com seu próprio ritmo. Assim, o estudante estará preparado para avançar em seus estudos com os próximos tópicos de programação de forma autônoma.

\subsection{Avaliação}

Para avaliar o desempenho dos estudantes neste projeto, serão realizadas simulações de edições anteriores de olimpíadas e competições promovidas e apoiadas pela SBC. Isso permitirá que os estudantes tenham um primeiro contato com o formato e nível de dificuldade das provas. Ao término da trilha, os estudantes serão submetidos a questionários e entrevistas com o objetivo de coletar informações sobre a rotina de estudos adotada. O aprimoramento do método e suas variações serão obtidos por meio da abordagem Design-Based Research (DBR), na qual a cada ciclo, as peculiaridades a serem melhoradas são identificadas e suas modificações propostas para os próximos ciclos \cite{Herrington2007}.

Até o presente momento, os alunos participaram da primeira fase da OBI, em junho de 2024, competindo na modalidade de programação nos níveis: Nível Júnior, Nível 1 e Nível 2. Além disso, os estudantes aprovados para a próxima fase continuam o treinamento para a segunda fase da OBI.

\section{Resultados e Discussões} \label{sec:resultados}

A avaliação dos efeitos do programa da OBI incluiu a análise do aumento de participantes e do progresso das habilidades em informática. Nesta parte, é feita uma análise dos dados recolhidos durante a primeira etapa da OBI. Os resultados são analisados considerando os objetivos estabelecidos, ressaltando as realizações e obstáculos enfrentados pela equipe e pelos alunos participantes.

\begin{figure}[htb]
    \parbox{6.5cm}{
        \includegraphics[width=0.9\linewidth]{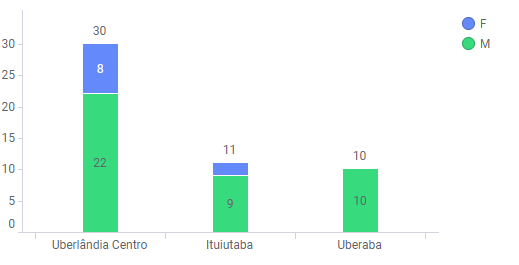}
         \caption{Participantes OBI - 2023}
         \label{fig:graph1}
    }
    \begin{minipage}{8.5cm}
         \includegraphics[width=0.99\linewidth]{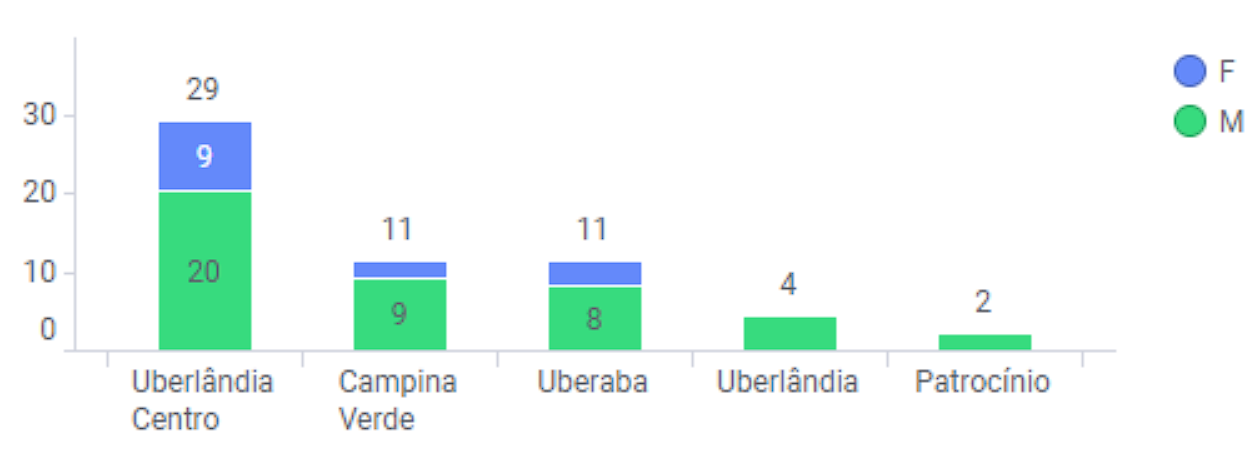}
         \caption{Participantes OBI - 2024}
         \label{fig:graph2}
    \end{minipage}
\end{figure}

Este estudo constatou, por meio da análise da participação de alunos do ensino médio e fundamental na OBI de 2024, demonstrado nas Figuras \ref{fig:graph1} e \ref{fig:graph2}, onde os estudantes estão separados por gênero, um aumento no número de unidades do IFTM participantes da OBI. Os campi Uberlândia Centro e Uberaba participaram da prova em 2023, tendo praticamente o mesmo número de alunos em 2024. O Campus Ituiutaba, que participou da prova em 2023, teve 21 inscritos em 2024 sendo 16 estudantes de ensino médio, porém os alunos não conseguiram realizar a prova por problemas com a infraestrutura do prédio.


Já Campina Verde também participou em 2023, porém não foi disponibilizado os dados da quantidade de participantes nessa edição, em 2024 onze estudantes fizeram a prova. É a primeira participação do Campus Patrocínio com estudantes do ensino médio e a primeira participação na OBI na história do Campus Uberlândia. Em 2024, dos 9 campi da instituição, 6 inscreveram os estudantes e dentre estes, apenas 1 não conseguiu aplicar a prova. Demostrando que em 2024, houve uma expansão de 2 campi, em comparação com 2023, reforçando o objetivo da fase inicial da pesquisa que é disseminar o conhecimento sobre programação e sobre a OBI. Estes dados podem sugerir um reflexo dos treinamentos oferecidos e indicam a disseminação do treinamento de programação e o conhecimento da OBI nos diferentes campi, no entanto, uma analise mais detalhada ainda será realizada. 


Considerando ainda os dados das figuras \ref{fig:graph1} e \ref{fig:graph2}, percebe-se que o número total de participantes femininas na competição obteve um leve crescimento em relação ao ano de 2024. Em  2023, 10 alunas participaram, enquanto no ano seguinte esse número subiu para 14. Apesar desse aumento em números absolutos, nota-se que ao comparar os resultados gerais de ambos os anos, o grupo feminino representa apenas 22,2\% do total de estudantes, evidenciando uma baixa participação feminina nesta competição e necessidade crescente de ações afirmativas.

\begin{figure}[htb!]
     \centering
     \includegraphics[width=0.6\linewidth]{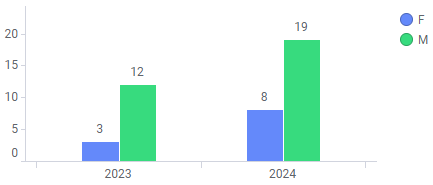}
     \caption{Quantitativo de participantes por gênero na modalidade P1 - Campus A}
     \label{fig:graph5}
 \end{figure}

A Figura \ref{fig:graph5} ilustra o quantitativo de participante, separados por gênero, considerando apenas os estudantes do Campus Uberlândia Centro, na modalidade de programação nível 1 (P1). Nessa amostra, especificamente, a participação das estudantes cresceu mais de 50\% em comparação ao ano de 2023, saltando de 3 alunas em 2023 para 8 alunas em 2024. Para essa turma, o treinamento foi realizado de maneira presencial e ministrado por uma instrutora mulher. 

\begin{figure}[htb]
    \parbox{7cm}{
        \includegraphics[width=0.95\linewidth]{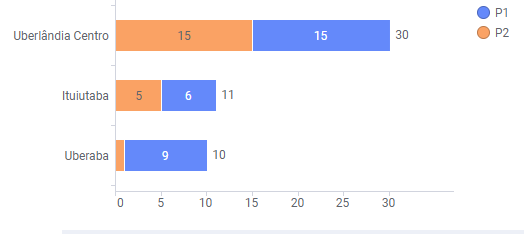}
         \caption{Participantes por nível - 2023}
         \label{fig:graph3}
    }
    
    \begin{minipage}{8cm}
         \includegraphics[width=0.98\linewidth]{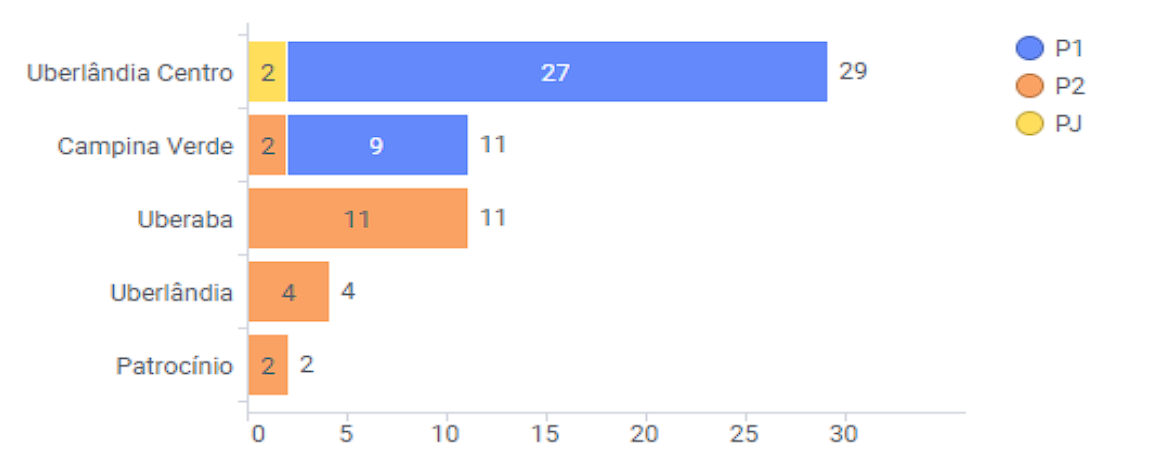}
         \caption{Participantes por nível - 2024}
         \label{fig:graph4}
    \end{minipage}
\end{figure}

Os dados apresentados pelas Figuras \ref{fig:graph3} e \ref{fig:graph4} representam o quantitativo de participantes na OBI por nível na modalidade programação em 2023 e 2024, respectivamente. Em relação a distribuição de participação entre as modalidades da OBI, além das respectivas faixas etárias, é observado que a modalidade P1 foi a mais recorrente em ambos os anos. No ano de 2023, os alunos presentes nesta modalidade representaram 58\% do total de participantes, englobando 30 alunos do total, enquanto em 2024 este valor aumentou para 36 alunos representando 60\% dos estudantes. Isto indica um padrão de interesse para esta categoria.


\begin{table}[htb!]
    \centering
    \caption{Quantitativo de participantes classificados para 2º fase}
    \label{tab: aprovados}
    \includegraphics[width=0.4\textwidth]{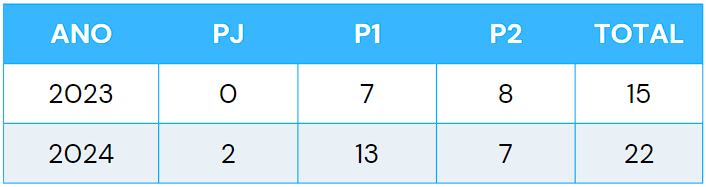}
\end{table}

No que concerne ao desempenho geral dos alunos na prova, observa-se, através da Tabela \ref{tab: aprovados}, que para o ano de 2024 houve um aumento no número de estudantes que se classificaram para a segunda fase da OBI. Destaca-se novamente alunos da modalidade P1, que obteve o maior número de classificados, resultando em um aumento de 8 alunos em comparação com 2023, evidenciando mais uma vez maiores interesses por parte desta modalidade. O treinamento continua sendo realizado, mesmo para os alunos que não se classificaram, pois o objetivo é continuar a preparação para a prova do próximo ano.

\begin{figure}[htb]
    \parbox{7.5cm}{
        \includegraphics[width=0.85\linewidth]{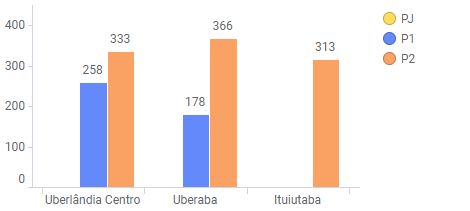}
         \caption{Participantes por nível - 2023}
         \label{fig:graph6}
    }
    \begin{minipage}{8cm}
         \includegraphics[width=0.95\linewidth]{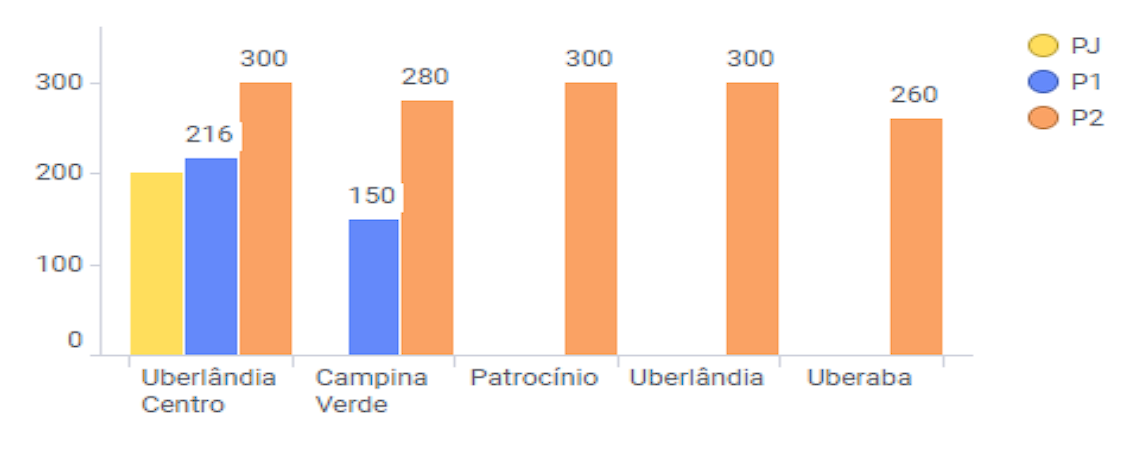}
         \caption{Participantes por nível - 2024}
         \label{fig:graph7}
    \end{minipage}
\end{figure}

Para concluir as análises deste estudo em andamento, observa-se que o desempenho médio das pontuações no ano de 2024 não apresentou valores significativos em comparação a 2023. Conforme demostrado pelas Figuras \ref{fig:graph6} e \ref{fig:graph7} que apresentam o valor médio obtido em cada um dos níveis, da fase 1 da OBI, considerando todos os participantes por campus. 


No entanto, é importante considerar que a maior parte dos estudantes participantes da modalidade P1 não possuíam nenhum conhecimento prévio em programação e que o IFTM esteve em greve durante o período do treinamento, resultando em menor participação dos alunos nas aulas. Em contrapartida, destaca-se o avanço no desempenho da modalidade PJ. Como mencionado anteriormente, houve uma participação inédita de alunos do nono ano, o que resultou na classificação de todos para a fase seguinte, inclusive da única aluna classificada, que pertence a esta modalidade.

\section{Próximos passos} \label{sec:passos}
Atualmente, a pesquisa encontra-se no primeiro ciclo de execução das turmas presenciais e remotas. Para os próximos passos, ao longo dos anos de 2024 a 2026, planeja-se realizar mais 5 ciclos, com duração de 5 meses cada, para os estudantes na modalidade remota vinculados ao Instituto Federal. Durante esse período, os estudantes participarão das provas da OBI e da Competição Feminina da OBI (CF-OBI), onde será realizada a coleta de dados dos resultados obtidos nas competições, seguida da análise e discussão desses resultados.

\section{Conclusão} \label{sec:conclusao}

Competições de programação ainda são uma modalidade de olimpíada científica pouco conhecida no Brasil e com baixo incentivo nas escolas. A partir da pesquisa realizada, os resultados preliminares demonstram que treinamento em programação é uma ferramenta relevante para o ensino de programação e promoção do pensamento computacional.

A partir da divulgação e parceria para viabilidade das turmas, houve um aumento no número de campus do IFTM que realizaram a prova da OBI, assim como maior número de alunos selecionados para a fase 2 da competição. Ressalta-se também crescimento na participação feminina na modalidade programação, nível 1, do campus Uberlândia Centro, indicando outro impacto positivo do treinamento.

Por se tratar de uma pesquisa em andamento, a análise de dados mais aprofundada, bem como a realização de novos ciclos de treinamento, são necessários para o refinamento e avaliação mais completa da metodologia proposta. 
Esta pesquisa visa destacar a importância de ações que promovam a inclusão e o acesso à tecnologia para todos, promover o pensamento computacional, incentivando a participação de estudantes em competições como a OBI e fomentando o desenvolvimento de habilidades computacionais essenciais para o mercado de trabalho e para a vida em sociedade.


\section{Agradecimentos}
Agradecemos a todos os membros da equipe de pesquisadores, professores e estudantes que contribuíram para a execução desta pesquisa. Manifestamos, ainda, nossa gratidão a todas instituições participantes do projeto. Reconhecemos, adicionalmente, o apoio da Faculdade de Computação (FACOM - UFU)  e a Coordenação de Aperfeiçoamento de Pessoal de Nível Superior - Brasil (CAPES).

\bibliographystyle{sbc}
\bibliography{referencia}

\end{document}